# Efficient Linear Programming Decoding of HDPC Codes

Alex Yufit, Asi Lifshitz and Yair Be'ery, *Senior Member IEEE*

Tel Aviv University, School of Electrical Engineering

Ramat Aviv 69978, ISRAEL

Email: yufitale@post.tau.ac.il, asilifsh@post.tau.ac.il, ybeery@eng.tau.ac.il

*Abstract* - We propose several improvements for Linear Programming (LP) decoding algorithms for High Density Parity Check (HDPC) codes. First, we use the automorphism groups of a code to create parity check matrix diversity and to generate valid cuts from redundant parity checks. Second, we propose an efficient mixed integer decoder utilizing the branch and bound method. We further enhance the proposed decoders by removing inactive constraints and by adapting the parity check matrix prior to decoding according to the channel observations. Based on simulation results the proposed decoders achieve near-ML performance with reasonable complexity.

*Index Terms* - linear programming, adaptive LP decoding, automorphism groups, branch and bound, LP relaxation, pseudocodewords, belief propagation, BCH codes.

## I. Introduction

Linear Programming (LP) decoding was proposed by Feldman in [1] for decoding binary linear codes. Feldman showed that ML decoding can be equivalently formulated as an LP optimization problem running on the codewords polytope. The codewords polytope is a convex hull of all the valid codewords, described by a set of constraints. The number of constraints depends exponentially on the code length; thus Feldman proposed a sub-optimal algorithm running on a relaxed codewords polytope. The relaxation of the codewords polytope is done as follows: each row of the parity check matrix is represented by a set of linear constraints. These constraints construct a local codeword polytope. The intersection of all the local codeword polytopes is a relaxed polytope, also called the fundamental polytope. The vertices of the fundamental polytope contain every codeword, but also some non-codeword pseudocodewords. The LP decoding process finds a vertex of the fundamental polytope maximizing the channel likelihood ratios function. One of the advantages of LP decoding is that the decoder has the desirable Maximum Likelihood (ML) certificate property; whenever it outputs an integral solution, it is guaranteed to be the ML codeword [1]. Another advantage is that the relaxed polytope can be tightened by including additional constraints, such as constraints based on Redundant Parity Checks (RPC) [2].

The work of [1] was focused on Low Density Parity Check (LDPC) codes. In the formulation of [1] the number of constraints per check node is exponential in the check node degree. LDPC codes are characterized by a sparse parity check matrix, which in turn leads to a relatively small set of LP constraints. High Density Parity Check (HDPC) codes are characterized by a dense parity check matrix, which leads to poor performance over the AWGN channel when using the decoder of [1] due to several reasons. First, the number of constraints depends exponentially on each parity check degree, which results in a very large linear problem. Second, each vertex of the fundamental polytope is the intersection of several hyper-planes, defined by the parity check constraints; thus increasing the number of constraints leads to increasing the number of pseudocodewords, deteriorating the performance of the decoder.

In [3], Yang et al. presented a compact LP relaxation which required less constraints compared to [1] to describe the relaxed polytope. Yang et al. proved in [3] that the new relaxation is equivalent to [1] by means of having the same projection on the codewords polytope; thus having the same performance. In the new formulation, the number of constraints is polynomial in the check node degree resulting in a considerable decrease of computational complexity. However, despite the complexity reduction, [3] cannot be applied to medium and long HDPC codes, since the number of constraints is still prohibitively large.

In [2] an Adaptive LP (ALP) decoder was presented, which can be applied to medium and long HDPC codes. The decoder of [2] achieves the same performance as the decoders of [1] and [3] by solving a set of compact LP problems instead of one problem with a huge number of constraints. ALP decoder iteratively converges to the same solution as [1] by adding constraints and dropping some old constraints. A randomized algorithm for finding row RPC cuts to tighten the relaxation is also presented in [2] as a way to improve the performance. Nevertheless, finding efficient methods of constructing RPC cuts was left open.

In [4] Draper et al. have suggested a mixed integer decoding for improving the performance of LP decoding over LDPC codes. By adding integer constraints to the ALP decoder of [2], the proposed mixed integer decoder can obtain ML decoding performance. However, the authors of [4] have claimed that their method cannot be applied to HDPC codes since the complexity becomes prohibitively large.

An LP decoding algorithm for HDPC codes with acceptable performance and complexity was proposed by Tanatmis et al. [5]. This new decoder, denoted as a New Separation Algorithm (NSA) performs significantly better then the LP decoder of [1], [2], [3] and [4] mainly due to a generation of new RPC cuts. The algorithm generating RPC cuts is based on Gaussian elimination of the parity check matrix. Recently, Tanatmis et al. have improved their decoder by finding additional algorithms for generating RPC cuts [6].

The contribution of this paper is in providing better adaptive LP decoding techniques for decoding HDPC codes. We present several methods for eliminating fractional pseudocodewords, while adding tolerable complexity. First, we propose to improve the performance of the LP decoder by running several decoders in parallel, while each one using a different representation of the parity check matrix. As an alternative, we propose using several representations of the parity check matrix to derive RPC cuts in order to tighten the relaxed polytope. The variety of parity check matrix representations are produced by columns' permutations taken from the Automorphism groups of the code. Second, a new variant of mixed integer decoding utilizing the branch and bound method is proposed. We further enhanced the proposed decoders by utilizing ordered statistics for adapting the parity check matrix according to the channel observations prior to decoding. We also propose removing all inactive constraints after each decoding iteration, which avoids the growth of the number of constraints and reduces the number of fractional pseudocodewords in the relaxed polytope. Our decoding algorithms adaptively seek for constraints which exclude fractional pseudocodewords from the feasible set. We achieve a near-ML performance by solving several compact LP problems, rather than solving a single growing problem.

The rest of the paper is organized as follows: We provide some preliminaries in Section II. In Section III we introduce an improved adaptive LP decoder which makes use of the automorphism groups. In Section IV we present an adaptive branch and bound decoding algorithm. In Section V we propose two further enhancements for the adaptive LP decoder: parity check matrix adaptation and removing inactive constraints. Simulation results are presented in section VI. Section VII concludes the paper.

## II. Preliminaries

A binary Linear code C of dimensions (n, k) is the space of all the binary words of length n which can be described by its parity check matrix H. A binary vector $\underline{x}$ of length n is a codeword iff $\underline{x}$ satisfies: $H\underline{x}=\underline{0}$ in GF(2). In this paper we consider transmitting over Additive White Gaussian Noise (AWGN) channel using BPSK modulation. Let $\underline{y}$ be the receiver input, then we can formulate the ML decoding problem in the following way:

$$\underline{x}_{ML} = \text{argmin } \{\sum_{i=1}^{n} c_i x_i \ \ s.t. \ \underline{c} \in C\} \tag{1}$$

where $c_i=\log(\Pr(y_i|x_i=0)/ \ \Pr(y_i|x_i=1))$ is the log likelihood ratio of the $i^{th}$ bit and $\underline{x}_{ML}$ is maximum likelihood codeword. The vector $\underline{c}=[c_1,…,c_n]$ is also called a cost vector.

Each check node j in the Tanner graph [7] of a code generates a local codeword polytope [1]. The local codeword polytope is defined by the following set of constraints:

$$\sum_{i \in N(j) \backslash S} x_i + \sum_{i \in S} (1-x_i) \geq 1, \ \ |S| \ is \ odd \tag{2}$$

where N(j) is a set of all the bit nodes connected to check node j and S is an odd subset of N(j). The constraints (2) are also known as Forbidden Set (FS) inequalities or FS constraints.

The decoder of [1] performs LP optimization over the fundamental polytope. When the LP solver returns an integer codeword it has the ML certificate property [1], which guarantees that the decoded word is the ML-codeword. However, when the solution contains fractional coefficients, the decoder returns an error. In [1] Feldman proposed adding RPC cuts in order to tighten the relaxed polytope and thus improve the performance. However, no efficient way of finding such RPCs was given.

The number of FS constraints required to describe the fundamental polytope in Feldman's LP decoder [1] is $\sum_{j=1}^{m} 2^{d_c(j)-1}$ where $d_c(j)$ is the order of the $j^{th}$ row of H, and m is the number of rows in H. Accordingly, the complexity of the decoder of [1] grows exponentially with the density of the parity check matrix; thus such a decoder is only applicable to LDPC codes.

The ALP decoder of [2] adaptively adds FS inequalities (2) and solves a set of compact LP problems rather than one large problem. In [2] it was shown that such an adaptive decoder converges to the same solution as the original LP decoder of [1], but with much less complexity. For each check node only a single constraint from the set S of (2) can generate a cut. Furthermore, the maximal number of iterations required is bounded by the code length. These bounds make adaptive LP decoding efficient even for HDPC codes. However, the performance of [2] is poor when applied to HDPC codes.

The NSA decoder proposed by Tanatmis et al. in [5] is also an adaptive LP decoder. In [5], additional auxiliary variables are used as indicators to detect the violated parity checks. Each indicator variable corresponds to a row of the parity check matrix and satisfies the following equality constraints:

$$H\underline{x} - 2\underline{z} = \underline{0} \tag{3}$$

$\underline{x}$ and $\underline{z}$ are the codeword bit variables and auxiliary variables respectively and $\underline{0}$ is the all-zeros vector. The initial LP problem of [5] contains only the constraints of (3). If the decoder output is integral, the ML certificate property guarantees that the output is the ML codeword. Otherwise, valid cuts are generated and added to the LP formulation to eliminate the fractional pseudocodeword. A modified LP problem is solved again repetitively until an integral solution is found or no more valid cuts could be generated.

 

Valid cuts are generated by either FS inequalities of (2) or by generating RPC cuts. RPC cuts are generated based on the observation that every fractional pseudocodeword is cut by an RPC whose support contains only one fractional index. Such RPCs are generated by Gaussian elimination of the parity check matrix as described in [5].

## III. LP Decoders utilizing Automorphism Groups

Automorphism group of a linear block code C, Aut(C), is a group of all the coordinates' permutations which send C into itself [8]. For example, the automorphism group of any cyclic code includes a cyclic shift permutation, as well as other permutations which depend on the internal structure of the code. For many widely used block codes the automorphism groups are well known. Automorphism groups of some primitive binary BCH codes are provided in [9].

Several BP decoders exploit the knowledge of the automorphism group of the code to generate alternative representations of the parity check matrix by permuting its columns with a permutation taken from Aut(C) [10], [11], [12]. In BP decoders, the choice of a parity check matrix has a crucial influence on the Tanner Graph [7] of the code and thus on the performance of the decoder. It is shown in [10], [11], [12] that a BP decoder running on several H matrices in parallel achieves significant performance improvement for HDPC codes, though increases the computational complexity.

We use the same approach with LP decoders based on the following: First, an LP decoder is also susceptible to the choice of the parity check matrix [1] i.e. different parity check matrices will have different relaxations of the codewords polytope and thus different pseudocodewords. A systematic way for a-priori choosing a representation of the parity check matrix that will yield the best decoding result for a given received word is still unknown; thus we propose to run several LP decoders in parallel, each one using a different parity check matrix representation. We denote this concept: *parity check matrix diversity.* Second, alternative representations of a parity check matrix can be exploited to derive new RPC cuts when using adaptive LP decoders [2], [5]. When an adaptive decoder is unable to generate any more valid cuts, alternative representations of the parity check matrix can provide additional RPCs to generate valid cuts in order to obtain a tighter relaxation. Each of the two approaches presented above can be used with any LP decoder ([1], [2], [3] and [5]). In the following we propose using the first approach with the NSA decoder of [5] and the second with the ALP decoder of [2].

*A. LP Decoder utilizing Parity Check Matrix Diversity*

The behavior of an adaptive LP decoder [2], [5] is difficult to analyze: At each stage, the relaxed polytope is modified due to the addition of new constraints and possibly disposal of old constraints [2]. These modifications lead to the emergence of some new pseudocodewords and the disappearance of some existing pseudocodewords. For each chosen parity check matrix, the adaptive LP decoder may converge to a fractional pseudocodeword minimizing the objective function in a vicinity of the received word and causing a decoding failure. Such a pseudocodeword may not exist in the relaxed polytope derived from a different parity check matrix.

The parity check matrix diversity is a powerful tool to reduce the likelihood of decoding failures caused by converging to pseudocodewords. Algorithm 1 is an improved version of the NSA decoder provided in [5], and its performance gain will be demonstrated in section VI. For each received word, the decoder tries to decode the word several times, each time with a different representation of the parity check matrix. Each new representation is generated by permuting the columns of H with a random permutation taken from Aut(C). The ML certificate property of the LP decoder is used as a stopping criterion when an integral solution is found. If none of the decoders returned a valid codeword, the algorithm returns a fractional solution with minimal Euclidian distance from the received word.

In the description of Algorithm 1 we used the following notation: $\underline{x}$ = argmin $\{\underline{c}'\underline{x}$ s.t. H} means run the NSA decoder which finds $\underline{x}$ that minimize $\underline{c}'\underline{x}$, subject to using H as a parity check matrix representation. The worst case complexity of Algorithm 1 is higher than the worst case complexity of the NSA decoder by a linear factor N, while resulting in a major performance enhancement. One can trade performance for complexity by controlling the maximal number of NSA decoding attempts, N.

**Input**:
        $\underline{r}$ – Received vector
        $\underline{c}$ – Cost vector
        H – Parity check matrix
        Aut(C) – Automorphism group of code C
        N - Maximal number of decoding attempts
**Output**:
        $\underline{x}_{opt}$ – Returned optimal solution

1. Init $\underline{x}_{opt} = \underline{0}$ , c= -$\underline{r}$.
2. For (i in range 1 to N) repeat
    2.1. Run the NSA decoder of [5]: $\underline{x}_i =$ argmin$\{\underline{c}^t\underline{x}$ s.t. H$\}$
    2.2. If $\underline{x}_i$ is integral then $\underline{x}_{opt}=\underline{x}_i$
        2.2.1. Permute the result: $\underline{x}_{opt}=\pi_1 \cdot \ldots \cdot \pi_{i-1}(\underline{x}_i)$
        2.2.2. Return $\underline{x}_{opt}$
        2.2.3. Terminate
    2.3. Choose randomly a permutation : $\pi_i \in$ Aut(C)
    2.4. Apply inverse permutation to the cost vector : $\underline{c}=\pi_i^{-1}(\underline{c})$
3. Permute the vectors: $\underline{x}_k=\pi_1 \cdot \ldots \cdot \pi_k(\underline{x}_k), \ k \in \{1,...,N\}$
4. Find a solution $\underline{x}_{opt} = $ argmin $\left\| \underline{x}_k - \underline{r} \right\|, k \in \{1,...,N\}$
5. Return $\underline{x}_{opt}$

**Algorithm 1 – NSA decoder of [5] with parity check matrix diversity**

## B. LP Decoder utilizing RPC Cuts based on Alternative Parity Check Matrix Representations

As was stated earlier, alternative parity check matrix representations can contribute to the search of new RPC valid cuts. Valid RPC cuts are generated from the rows of the permuted H. Such an approach achieves tighter relaxation of the codeword polytope and results in a superior performance. This approach is demonstrated in Algorithm 2 based on the ALP decoder of [2].

The first stage of the proposed algorithm is to decode the received word using the ALP decoder of [2]. If the result is integral, the ML certificate property holds and the optimal solution is returned. Otherwise, a new representation of the parity check matrix is generated. Such a new representation is achieved by permuting the columns of H by a random permutation taken from Aut(C). The ALP decoder is used repetitively, each time with a different representation of H. The decoder continuous until an integral

 

solution is found or a maximal number of decoding attempts is reached. The major difference between the current approach and the approach presented in sub-section A is that the current approach preserves constraints generated in pervious decoding attempts.

The worst case complexity of Algorithm 2 is higher by factor N than the worst case complexity of the ALP decoder of [2]. The parameter N is chosen as a trade-off between the desirable performance and complexity.

---

**Input**:

  $\underline{r}$ – Received vector

  $\underline{c}$ – Cost vector

  H – Parity check matrix

  Aut(C) – Automorphism group of code C

  N - Maximal number of decoding attempts

  CS – Set of initial constraints

**Output**:

  $\underline{x}_{opt}$ – Returned optimal solution

1. Init c= -$\underline{r}$ , CS ={}
2. For (i in range 1 to N) repeat
    2.1. Run the ALP decoder of [2]: $\underline{x}$ = argmin {$\underline{c}^t\underline{x}$ s.t. H} with constraints set CS .
    2.2. Add all new generated constraints in 2.1 to the set CS.
    2.3. If $\underline{x}$ is Integral then goto 3.
    2.4. Choose a random permutation : $\pi_i \in$ Aut(C)
    2.5. Permute the columns of the parity check matrix H:  H= $\pi_i$(H)
3. Return $\underline{x}_{opt}=\underline{x}$ .

**Algorithm 2 – improved ALP decoder of [2] using alternative H representations**

---

## IV. A Branch and Bound LP-based Adaptive Decoder

The branch-and-bound method, in the context of linear programming decoding, was first proposed in [13], as a multistage LP decoding of LDPC codes. The proposed decoder is a suboptimal LP decoder, for which deeper depths lead to a better performance at the cost of an increased complexity. It can be applied to both LDPC and HDPC codes, does not use mixed integer programming, and requires less computations compared to [4]. It is able to refine the results of [5] by systematically eliminating the most unreliable fractional elements towards finding the ML-codeword

The proposed decoder initially tries to find the ML-codeword by calling the NSA decoder. If a fractional pseudocodeword was returned and no new valid cuts could be found, our decoder recursively constructs two new LP problems and solves them using [5]. The decoder adopts a depth-first tree searching approach of limited depth. At each node a different LP problem is solved. The root of the tree (depth equals zero) is the fractional solution (pseudocodeword) obtained by solving the original problem using [5]. A node at depth i has i new equality constraints, and has two children at depth i+1. The proposed algorithm is described below as Algorithm 3 and makes use of procedure BB_decode presented just after it.

---

**Input:**

        $\underline{c}$ - Cost vector

        H - Parity check matrix

        $BB_{depth}$ - Maximal depth

        $B_{upper-bound}$ - Objective value upper bound

**Output:**

        $\underline{x}_{opt}$ - Optimal solution

1. Run the NSA decoder of [5]: $\underline{x} = $ argmin$\{$ $\underline{c}^t\underline{x}$ s.t. H$\}$
2. If the output of [5] is integral then

        return $x_{opt} = \underline{x}$ ,Terminate
3. Find the fractional $x_i$ with the smallest absolute LLR and construct 2 LP problems:

    a. ORG_BB_zero: Original problem with a constraint $x_i$=0

    b. ORG_BB_one: Original problem with a constraint $x_i$=1
4. Set current_depth = 1, $B_{upper-bound}$ = MAXINT and $x_{opt}$ = $\underline{0}$
5. Call the procedure BB_decode:

    ($\underline{x}_{opt}$, $B_{upper-bound}$, current_depth) = BB_decode (ORG_BB_zero, current_depth, $BB_{depth}$, $B_{upper-bound}$, $x_{opt}$)
6. set current_depth=1

    ($\underline{x}_{opt}$, $B_{upper-bound}$, current_depth) = BB_decode (ORG_BB_one, current_depth, $BB_{depth}$, $B_{upper-bound}$, $x_{opt}$)

**Algorithm 3: Branch and Bound separation decoder**

---

$(\underline{x}_{opt}, B_{upper\text{-}bound}, current\_depth) = \textbf{BB\_decode}\ (LP\_problem, current\_depth, BB_{depth}, B_{upper\text{-}bound}, x_{opt})$

**Input:**

        LP_problem – LP problem

        current_depth – Current tree depth

        $BB_{depth}$ - Maximal depth

        $B_{upper\text{-}bound}$ - Objective value upper bound

**Output:**

        $\underline{x}_{opt}$ - Optimal solution

        $B_{upper\text{-}bound}$ - Objective value upper bound

        current_depth – Current tree depth

1. If current_depth > $BB_{depth}$ return $(\underline{x}_{opt}, B_{upper\text{-}bound}, current\_depth)$
2. Run the NSA decoder of [5] : $\underline{x} = \text{argmin}\{\ \underline{c}^t\underline{x}\ \text{s.t. H}\}$, denote the objective value as c*
3. If no feasible solution is found return $(\underline{x}_{opt}, B_{upper\text{-}bound}, current\_depth)$
4. If the solution is integral

        if c* < $B_{upper\text{-}bound}$ then

                $B_{upper\text{-}bound} = \underline{c}*$

                $\underline{x}_{opt} = \underline{x}$

        endif

        return $(\underline{x}_{opt}, B_{upper\text{-}bound}, current\_depth)$

    else

        if c* < $B_{upper\text{-}bound}$

                current_depth = current_depth + 1

                Find the fractional $x_i$ with the smallest absolute LLR and construct two problems:

                    • BB_zero: Current problem with a constraint $x_i$=0

                    • BB_one: Current problem with a constraint $x_i$=1

                Solve BB_zero and BB_one by calling the subroutine BB_decode

        endif

    endif

**Procedure BB_decode**

The decoders of [1], [2] and [5] make use of the ML-certificate property as a stopping criterion. The new equality constraints may lead to infeasible solutions or integral solutions which are not the ML-codewords. In order to maintain the ML-certificate property, our decoder has three pruning possibilities:

1) Pruning by infeasibility: If equality constraints violate the parity check matrix constraints, adding more constraints (deeper search) will not make the solution feasible.

2) Pruning by bound: The smallest integral solution found so far is stored as an incumbent. The optimal cost value of the incumbent is stored as $B_{upper\text{-}bound}$. If the cost value of a node is bigger than $B_{\text{upper-bound}}$, this node will not be fathomed; this is because each of its two children has an additional constraint, which in turn may only increase the cost.

3) Pruning by integrality: An integral solution at node i is a stopping criteria, since fathoming the node may only increase the cost.

The complexity of our algorithm is upper bounded by the complexity of solving $2^{D_p}$ LP problems using the NSA decoder, where Dp is the maximal allowed depth. It should be noted that each such problem does not use the RPC constraints of its predecessor; thus a compact problem is being solved at each node.

*Example of branch and bound separation algorithm:*

Let H be the parity check matrix of the Hamming [8,4,4] code

$$H = \begin{bmatrix} 1 & 1 & 1 & 1 & 1 & 1 & 1 & 1 \\ 0 & 1 & 0 & 1 & 0 & 1 & 0 & 1 \\ 0 & 0 & 1 & 1 & 0 & 0 & 1 & 1 \\ 0 & 0 & 0 & 0 & 1 & 1 & 1 & 1 \end{bmatrix}$$

Let r be an arbitrary received vector (LLR):

$$r = \begin{bmatrix} 0.798337, 1.421758, -1.240177, -0.771128, -1.745193, 0.554868, 0.983861, -0.404989 \end{bmatrix}$$

Figure 1 represents the algorithm steps for maximal depth equals 3. Edges enumeration represents the decoder steps. Vertices enumeration represents the additional binary constraints at a given depth.

One can observe that the tree is pruned twice: After step 5 and after step 6 due to pruning by bound. The final solution is found after 6 iterations of the NSA decoder. A different search algorithm could converge to the ML solution after two steps if the first search was for $x_8=1$. Different search techniques may improve the complexity, but have no affect on the decoder performance.

## V. Further Enhancements

### A. Adapting the Parity Check Matrix According to the Channel Observations

The method of adapting the parity check matrix was proposed for the sum product algorithm by Jiang and Narayanan [14]. We utilize this technique in LP decoding. In [14] matrix adaptation is performed before each decoding iteration. Due to the nature of LP decoders, for which no LLR refinement is obtained, the matrix adaption is performed only once prior to decoding. We perform Gaussian elimination, such that a sparse sub matrix that corresponds to the less reliable bits is created prior to decoding. We show in the next section that adapting the parity check matrix prior to decoding can significantly improve both the performance and the complexity. The later is achieved due to a faster convergence of the decoding algorithm. In [2], Taghavi et al. have proven that a fractional solution of the LP decoder requires the existence of a cycle in the Tanner graph over all fractional-valued nodes. Less reliable bits are more prone to become fractional due to their low weight in the objective function. By employing matrix adaptation, these cycles are pruned, and better performance is obtained.

### B. Removing Inactive Constraints

The performance and complexity of any adaptive LP decoder are strongly related to the number of constraints in the LP problem. An LP decoder based on the Simplex algorithm searches for an optimal solution over the vertices of the relaxed polytope. The number of codewords vertices of any relaxed polytope is constant, while adding new constraints increases the total number of vertices. Obviously, all the new added vertices are non-codewords fractional pseudocodewords. On one hand adding new constraints to the problem makes the relaxed polytope tighter, but on the other hand increases the number of non-codewords pseudocodewords, leading to more decoding failures. Furthermore, having a large number of constraints increases the decoding complexity, especially for long codes. In [2], Taghavi et al. proposed a way of detecting and disposing unused constraints without loss of performance. In their decoder each row of the parity check matrix can contribute at least one constraint, while the rest of the constraints are disposed. However, such a technique cannot be applied to the constraints generated from RPCs. The NSA decoder does not supply a way of removing old constraints, thus making the LP problem grow at each iteration.

The following propositions 2 and 3 show that the convergence of the adaptive LP decoder is influenced by a small number of active constraints. Active constraints are the constraints for which strict equality holds. In contrast, the nature of the adaptive LP decoder is to accumulate inactive constraints from previous iterations which do not contribute to the convergence but only increase the number of pseudocodewords. We propose disposing all inactive constraints at each iteration. The number of remaining constraints is bounded by the LP problem dimensions. Practically, inactive constraints can be easily detected by examining the slack variables returned by the LP solver.

*Proposition 2: Removing all the inactive constraints from the LP problem does not affect the optimal solution.* (Proof appears in Appendix I).

*Proposition 3: Adaptive LP decoder will never return the same solution twice when all the inactive constraints are removed at each iteration.*

Proof: We will prove that the objective function of the adaptive decoder increases monotonically. At each decoding iteration there are two possibilities: I. No valid cuts were found and II. Valid cuts were found and added as constraints to the LP problem. Case I represents a stopping criterion; thus we will concentrate on case II. In case II we add some new constraints given by valid cuts and remove some old inactive constraints. Based on Proposition 2 hadn't we added new constraints but only removed inactive constraints would not affect the LP solution. But because we added new constraints which cut the current LP solution, the objective function must increase in the next iteration. Repeating this at each iteration proves the statement.

QED.

It should be emphasized that the enhanced decoder doesn't converge to the same solution since it runs on a different relaxed polytopes. The enhanced decoder has fewer constraints; thus naturally its polytope will have less pseudocodewords. The difference between the presented approach and the one proposed in [2] is in the following: In the current approach all the inactive constraints are removed (including those generated from RPCs by cut generating Algorithm 2 of [5]). In contrast, in [2] only the constraints generated by the same row of parity check matrix are disposed, while the constraints generated from alternative parity check matrix representations and RPCs cannot be disposed. The equality constraints (3) cannot be disposed since they are always active by definition.

## VI. Simulations Results

The performance of the proposed decoders was tested by computer simulation. In the following, we present results for BCH[63,39,9] and BCH[63,36,11] codes. We choose three decoders, each one comprising of a specific combination of the presented techniques and enhancements.

    A.  Decoder A: Algorithm 1 with parity check matrix adaptation and removing inactive constraints.

    B.  Decoder B:  Algorithm 3 with parity check matrix adaptation.

    C.  Decoder C: NSA decoder with parity check matrix adaptation and removing inactive constraints.

In Decoder A we use diversity of order 5 as a tradeoff between performance and complexity. Regardless of the choice of the code, higher diversity order achieves only a minor performance improvement at a cost of linear increase of complexity. For Decoder B, the maximal depth, Dp, required to achieve near-ML performance varies for each code. In order to have approximately a 0.1dB gap from the ML curve, Dp=4 and Dp=6 are chosen for BCH[63,39,9] and BCH[63,36,11], respectively. Decoder C is an improved NSA decoder with both superior performance and reduced complexity. Simulation results for the mentioned decoders appear in Figs. 2 and 3. As a benchmark, the curves of NSA decoder and ML are plotted. Complexity is estimated as an average run-time of a decoded word, while all decoders run on the same machine. All simulations were taken on the same Linux machine (CentOS release 4.5, 8G RAM), using the same LP solver (CPLEX 10.2).

Decoder A achieves 0.3 to 0.4dB performance gain in frame error rate compared to the NSA decoder with almost the same complexity, for both BCH[63,39,9] and BCH[63,36,11] codes. Decoder B achieves 0.5 to 0.7dB gain, for BCH[63,36,11] and 0.3 to 0.5dB gain for BCH[63,39,9]. Decoder C achieves 0.2 to 0.3dB gain with half of the complexity the NSA decoder. Due to the ML certificate property, the complexity of Decoders A and B decreases sharply with the growth of SNR. At high SNR the complexity of all the presented algorithms is slightly inferior compares to the NSA decoder, since parity check matrix adaptation requires at least one Gaussian elimination of the parity check matrix.

## VII. Conclusions

Several improved adaptive LP decoders efficiently operating on HDPC codes were presented. A new LP decoder based on parity check matrix diversity was presented. This decoder outperforms the NSA decoder by more than 0.4dB. A new branch and bound decoder was proposed, achieving performance within 0.1dB of the optimal ML decoder. Two further enhancements were proposed: adaptation of the parity check matrix prior to decoding and disposing inactive constraints. Both enhancements were shown to achieve both performance gain and complexity reduction. The benefit of each algorithm was examined, as well as its performance versus its complexity. Based on the presented simulation results, a decoder that satisfies a given performance and latency requirements can be chosen among the presented techniques.

## Appendix I

*Proof of proposition 2:*

It is enough to prove that removing a single inactive constraint does not affect the LP solution. Denote $\underline{c}=(c_1,\ldots,c_n)$ as the cost vector, $\underline{x}=(x_1,\ldots,x_n)$ as the variables vector of the LP problem and A as an m by n constraints matrix which rows are $\{a_i|i=1 \text{ to } m\}$, for which each row represents an inequality constraint. Using the above notations, the LP problem is formulated as follows:

$$min \ \underline{c}^t\underline{x} \ \text{s.t} \ Ax \leq \underline{b} \tag{A1}$$

Without loss of generality, suppose $\underline{P}$ is the optimal solution of problem (A1), and the row $\underline{a}_1$ corresponds to an inactive constraint i.e.

$$\underline{a}_1\underline{x} < \underline{b} \tag{A2}$$

Hyper-plane (A2) divides the space into two sub-spaces as shown in Fig. 4. The sub-space on the right side of the hyper-plane is infeasible. After removing inactive constraint (A2) from the LP problem and solving the new problem, new optimal solution is found. Let $\underline{Q}$ be the optimal solution after removing (A2). We have to prove that $\underline{P}=\underline{Q}$. First Note that:

$$\underline{c}^t\underline{Q} < \underline{c}^t\underline{P} \tag{A3}$$

Also note that $\underline{Q}$ must lie on the right side of the hyper-plane (A1) i.e. in the infeasible part because otherwise $\underline{Q}$ would have been the optimal solution of (A1). Let l be the line connecting $\underline{P}$ and $\underline{Q}$ and let $\underline{R}$ be the crossing point of l and the hyper-plane (A2) such that

$$\underline{R} = t \ \underline{P} + (1\text{-}t)\underline{Q} \ , \ t \in (0,1) \tag{A4}$$

Note that $\underline{R}$ is a feasible point of (A1). After substituting (A3) into (A4) we get

$$\underline{c}^t\underline{R} = \underline{c}^t \ [t \ \underline{P} + (1\text{-}t)\underline{Q}] < \underline{c}^t\underline{P} \tag{A5}$$

This is a contradiction of the assumption that $\underline{P}$ is the optimal solution of (A1) and of the hyper-plane (A2) being an inactive constraint of (A1). Thus P=Q, i.e removing any inactive constraint does not affect the optimal solution.

*QED.*

# References

[1] J. Feldman, M. J. Wainwright, and D. R. Karger, "Using linear programming to decode binary linear codes," IEEE Trans. Inf. Theory, vol.51, no. 1, pp. 954–972, Jan. 2005.

[2] M.H. Taghavi, Amin Shokrollahi and P.H. Seigel, "Efficient Implementation of Linear Programming Decoding", arXiv:0902.0657v1 [cs.IT].

[3] K. Yang, X. Wang and J. Feldman, "A New Linear Programming Approach to Decoding Linear Block Codes", IEEE Transactions on Information Theory, vol.54, no. 3, pp. 1061-1072, Mar. 2008.

[4] Draper, S.C.; Yedidia, J.S. Wang, Y. "ML Decoding via Mixed-Integer Adaptive Linear Programming", IEEE International Symposium on Information Theory (ISIT), June 2007.

[5] A. Tanatmis, S. Ruzika, H.W. Hamacher, M. Punekar, F. Kienle and N. When, "A separation algorithm for improved LP decoding of linear block codes", arXiv:0812.2559v1 [cs.IT]

[6] A. Tanatmis, S. Ruzika, H.W. Hamacher, M. Punekar, F. Kienle and N. When, "Valid Inequlities for Binary Linear Codes" 2009 IEEE International Symposium of Information Theory, July 2009 ,Seoul, Korea.

[7] N. Wiberg, "Codes and Decoding on general graphs" Ph.D ,Linkoping University, Linkoping, Sweden 1996.

[8] F. J. MacWilliams and N. J. A. Sloane, "The theory of error correcting codes", North Holland Pub. Co., Amsterdam, Holland, 1977.

[9] Chung Chin Lu and Loyd R. Welch, "On automorphism Groups of Binary Primitive BCH codes" in Proc. IEEE International Synp. On Information Theory, Trondheim, Norway, 1994, p.51.

[10] T. R. Halford and K. M. Chugg, "Random redundant iterative soft-in soft-out decoding," IEEE Trans. Commun., vol. 56, no. 4, pp. 513-517, Apr. 2008.

[11] T. Hehn, J.B. Huber, S. Leander and O. Milenkovich , "Multiple-Bases Belief-Propagation for Decoding of Short Block Codes", ISIT 2007, Nice, France ,2007.

[12] I. Dimnik and Y. Be'ery, "Improved Random Redundant Iterative HDPC Decoding", IEEE Trans. Comm. Vol 57. No.7 pp.1982-1985, July 2009.

[13] K. Yang, X. Wang, and J. Feldman, "Non-linear programming approaches to decoding low-density parity check codes," IEEE Journal on Selected Areas in Communications, vol.24, no.8, pp. 1603-1613, Aug.2006.

[14] J. Jiang and K. R. Narayanan, "Iterative soft decision decoding of Reed-Solomon codes based on adaptive parity check matrices," IEEE Trans. Inf. Theory, vol. 52, no. 8, pp. 3746-3756, Jan. 2006.

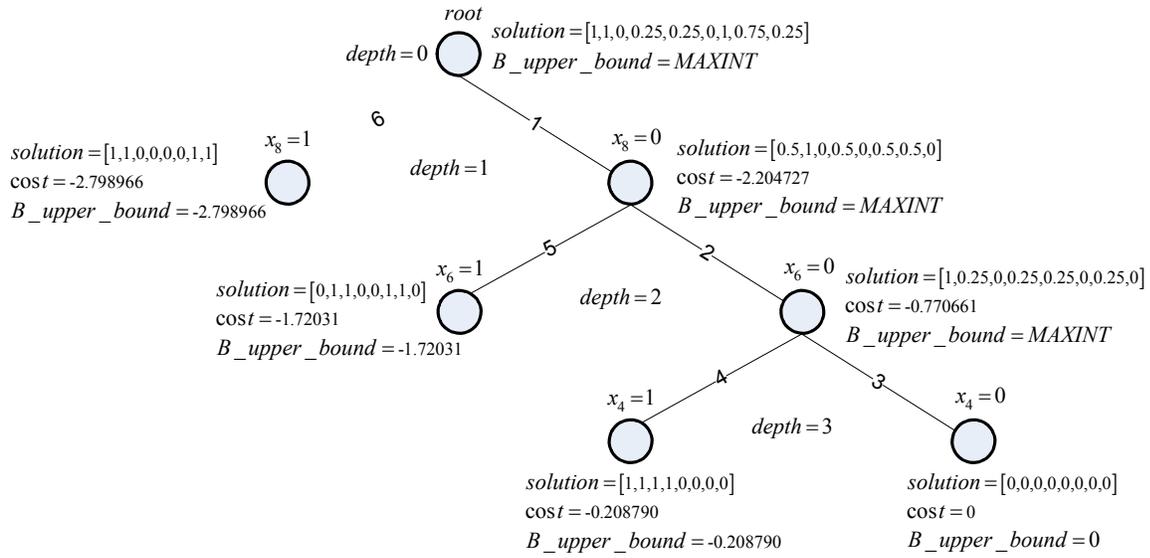

**Fig. 1.** Example of the Branch and Bound seperation algorithm

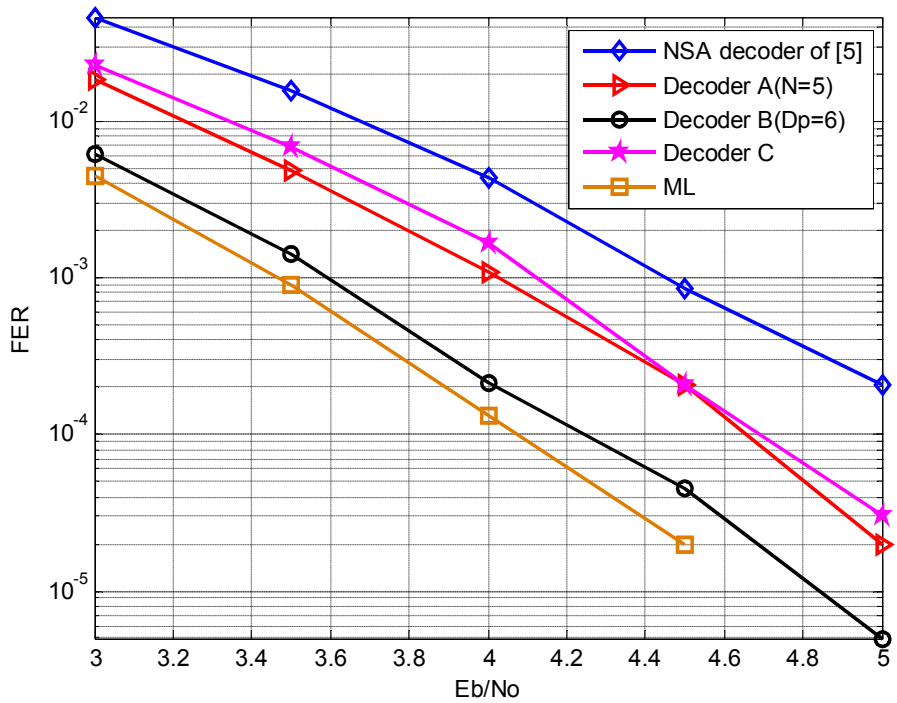

(a)  Frame Error Rate for BCH[63,36,11]

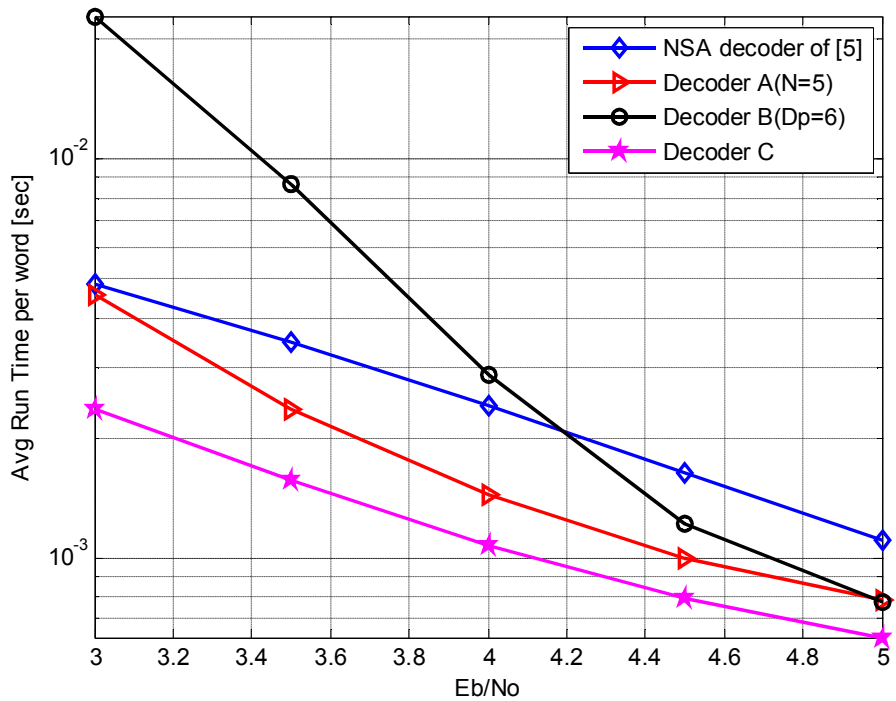

(b) Complexity for BCH[63,36,11]

**Fig. 2.**  Simulation results for BCH[63,36,11]

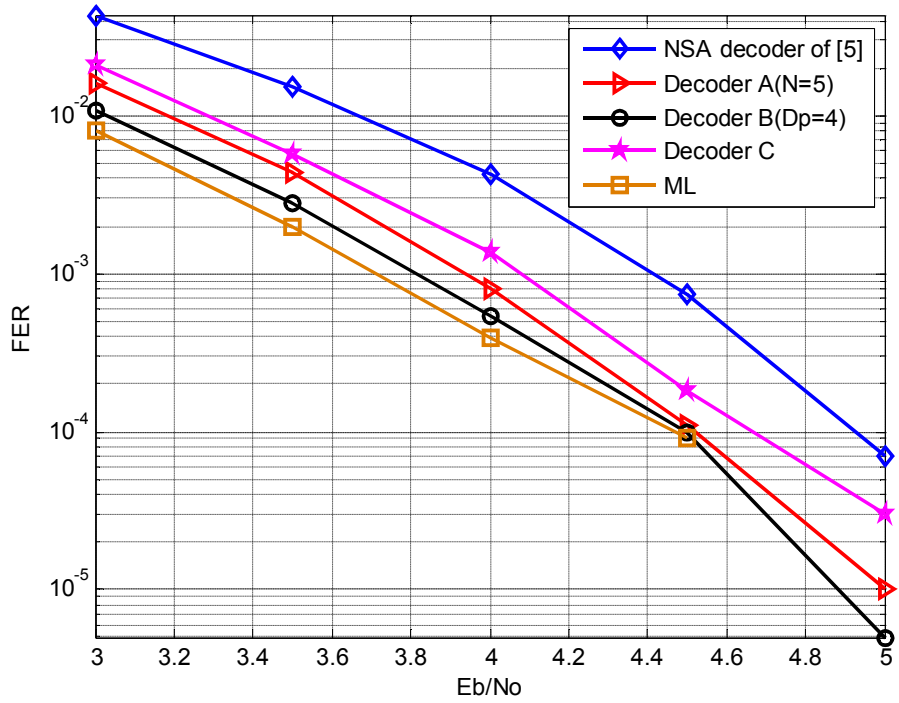

(a) Frame Error Rate for BCH[63,39,9]

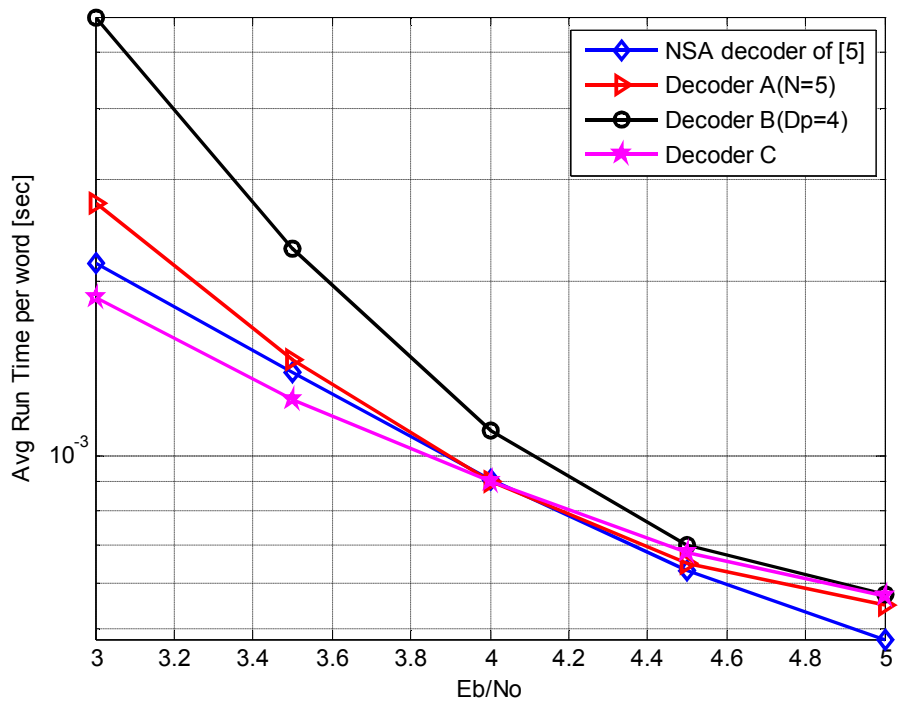

(b) Complexity for BCH[63,39,9]

**Fig. 3.** Simulation results for BCH[63,39,9]

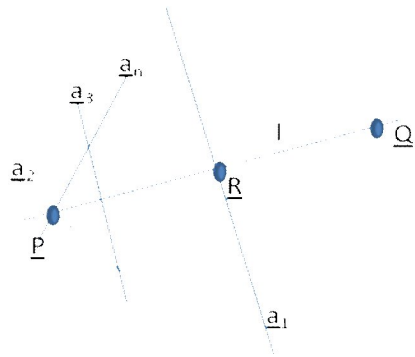

**Fig. 4**. Geometrical example of removing an inactive constraint